\documentclass[12pt,notitlepage]{article}
\usepackage{amsmath,amssymb}
\usepackage{graphicx}
\usepackage{cite}

\usepackage{color}

\def\ignore#1{{}}

\newcounter{sxn}

\newcounter{axn}

\newdimen\mybaselineskip
\newcommand{\beeq}{\begin{equation}}
\newcommand{\eneq}{\end{equation}}
\newcommand{\beqn}{\begin{eqnarray}}
\newcommand{\eeqn}{\end{eqnarray}}

\newcommand{\ba}{\begin{array}}
\newcommand{\ea}{\end{array}}

\newcommand{\be}{\begin{equation}}
\newcommand{\ee}{\end{equation}}
\newcommand{\bea}{\begin{eqnarray}}
\newcommand{\eea}{\end{eqnarray}}
\newcommand{\beal}{\setcounter{letter}{1} \begin{eqnarray}}
\newcommand{\eeal}{\addtocounter{equation}{1} \end{eqnarray}}

\newcommand{\larrow}{\,\,\,\,\hbox to 30pt{\rightarrowfill}
\,\,\,\,}
\newcommand{\slarrow}{\,\,\,\hbox to 20pt{\rightarrowfill}
\,\,\,}

\def\la{\raise.16ex\hbox{$\langle$}\lower.16ex\hbox{}  }
\def\ra{\, \raise.16ex\hbox{$\rangle$}\lower.16ex\hbox{} }

\def\psibar{ \psi \kern-.65em\raise.6em\hbox{$-$} \lower.6em\hbox{} }
\def\psibarb{ \psi \kern-.65em\raise.6em\hbox{$-$}  }

\begin{document}

\thispagestyle{empty}





\begin{center}  
{\LARGE \bf   Scalar Perturbations and Stability of a Loop Quantum Corrected Kruskal Black Hole}

\vspace{1cm}

{\bf  Ramin G.~Daghigh$^{1}$, Michael D.\ Green$^2$, Gabor Kunstatter$^3$,}
\end{center}

\centerline{\small \it $^1$ Natural Sciences Department, Metropolitan State University, Saint Paul, Minnesota, USA 55106}
\vskip 0 cm
\centerline{} 

\centerline{\small \it $^2$ Mathematics and Statistics Department, Metropolitan State University, Saint Paul, Minnesota, USA 55106}
\vskip 0 cm
\centerline{}

\centerline{\small \it $^3$ Physics Department, University of Winnipeg, Winnipeg, MB Canada R3B 2E9}
\vskip 0 cm
\centerline{} 

\vspace{1cm}

\begin{abstract}
We investigate the massless scalar field perturbations of a new loop quantum gravity motivated regular black hole proposed by Ashtekar {\it et al.} in [Phys.Rev.Lett. 121, 241301 (2018), Phys.Rev.D 98, 126003 (2018)].  The spacetime of this black hole is distinguished by its asymptotic properties: in Schwarzschild coordinates one of the metric functions diverges as $r\to \infty$ even though the spacetime is asymptotically flat. We show that despite this unusual asymptotic behavior, the quasinormal mode potential is well defined everywhere when Schwarzschild coordinates are used.  
We propose a useful approximate form of the metric, which allows us to produce quasinormal mode frequencies and ringdown waveforms to high accuracy with manageable computation times.
Our results indicate that this black hole model is stable against massless scalar field perturbations.  We show that, compared to the Schwarzschild black hole, this black hole oscillates with higher frequency and less damping. We also observe a qualitative difference in the power-law tail of the ringdown waveform between this black hole model and the Schwarzschild black hole.  This suggests the quantum corrections affect the behavior of the waves at large distances from the black hole.
	
\end{abstract}

\newpage

\section{Introduction}

The Nobel prize winning singularity theorem of Penrose\cite{Penrose1965} proves that general relativity invariably leads to singularity formation inside black hole event horizons, thereby signaling its own demise. It is commonly believed that the singularity will be resolved by the ultimate microscopic theory, presumably a version of quantum gravity, that describes the final stage of collapse. Since we are ostensibly quite far from a complete theory of quantum gravity that can accurately describe such a process, it is useful to utilize toy models in order to study the structure of  the presumable non-singular complete spacetime associated with the formation and evaporation of regular black holes (RBHs).  Although deviations from the Schwarzschild solution in the spherically symmetric case might normally only be significant near the Planck scale, it may in fact be that new physics enters at a different length scale, one that is accessible to astrophysical observations. This  possibility is given more credence by consideration of the so-called information loss paradox\cite{InfoLoss}. According to Page's arguments\cite{Page1993} any mechanism that enables information to emerge from an evaporating black hole horizon must become significant about half way through the evaporation process (the Page time) at which point the horizon radius can be very large. It is in fact millions of kilometers for galactic black holes. There are examples in the literature of dynamical theories  in which the new length scale associated with singularity resolution is macroscopic. See for example \cite{reg9} and \cite{Teukolsky}.

Many models of RBH spacetimes have been studied over the years.  See some of the Schwarzschild-like RBHs in \cite{reg1, PoissonIsrael, reg2,reg3,reg4, reg5, reg6, reg7, reg10,reg11, reg12, reg14}. In many cases, the spacetime metrics are not derived from any underlying microscopic theory and are  not closely connected to a potential theory of quantum gravity. Two notable exceptions, in the context of loop quantum gravity (LQG),  appear in \cite{pk09}\footnote{A RBH spacetime similar to that in \cite{pk09} is also derived by Modesto\cite{modesto06}.} by Peltola and Kunstatter and more recently in \cite{Ashtekar1, Ashtekar2} by Ashtekar, Olmedo and Singh in which complete regular static black hole spacetimes are derived as solutions to an effective theory motivated by LQG. For brevity we call the former the PK (Peltola-Kunstatter) black hole and the latter the AOS (Ashtekar-Olmedo-Singh)  black hole. An interesting feature of both these RBHs is that the singularity is in effect avoided  by the removal of $r=0$ from the spacetime and its replacement by a minimum area\footnote{LQG naturally leads to such a minimum area element. }  whose value is ultimately determined by the microscopic theory. As in the case of the Schwarzschild region, spatial slices that extend to spatial infinity describe an Einstein-Rosen wormhole with a minimum (throat) radius that changes with Kruskal time. For both the PK and AOS black holes, the throat radius shrinks to its minimal value before re-expanding. In the case of the PK black hole the throat radius re-expands to infinity producing in the future interior a single Kasner-type cosmological spacetime. For the AOS black hole the re-expansion proceeds until the formation of a second horizon whose radius is the same as that of the first horizon. The solution can then be analytically continued to produce a new (time-reversed) asymptotic region. The resulting Penrose diagram consists of an infinite tower of horizons and asymptotic spacetimes.  

In a previous paper \cite{DGMK}, three of the current authors studied the effects of singularity resolution on the response to perturbations of the PK black hole, including the quasinormal mode (QNM) spectrum and the ringdown waveform. Such calculations are potentially relevant to gravitational wave observations as well as to determining the stability\cite{RW} of the black hole solutions under consideration. 

The purpose of the present paper is to do a similar study of the regular spacetime of the AOS black hole\cite{Ashtekar1, Ashtekar2}. This particular spacetime is noteworthy not only because of its close connection to an underlying microscopic theory, but also because of its interesting analytic structure and global properties. 
In Schwarzschild-like coordinates, the $g_{00}$ metric component has a pre-factor of the form $(r/r_H)^\epsilon$, where $r_H$ is the horizon radius and $\epsilon \ll 1$ is a dimensionless parameter that derives from the microscopic theory.  
Because of this pre-factor, the $g_{00}$ metric component diverges to infinity as $r\rightarrow \infty$.
The AOS metric is nonetheless asymptotically flat, as can be verified by calculating the relevant curvature invariants which approach zero at infinity, albeit at a slower rate than for other quantum corrected black holes \footnote{We note that perturbations of other RBH models are investigated in, for example, \cite{QNMLQGbh0, QNMreg1, QNMreg2, QNMreg3,QNMreg4, QNMreg5, QNMreg6, QNMreg7, QNMreg8, QNMreg9, QNMLQGbh1,QNMreg10,QNMreg11, QNMreg12, QNMreg13, QNMreg14,QNMLQGbh2, QNMreg15}.}. Although the Schwarzschild form of the metric does not approach Minkowski spacetime at spatial infinity, Ashtekar and Olmeda\cite{Ashtekar3} showed that a time dependent change of coordinates puts the metric in a form that does approach Minkowski at infinity. In the new coordinates, the time is not the Killing time and the existence of a time-like Killing vector in the exterior is not manifest. It is of interest to determine how this asymptotic structure affects observables such as QNMs and waveforms associated with macroscopic black holes.


We structure the paper as follows. In Sec.\ \ref{Sec:WE}, we set up the problem by introducing the QNM wave equation. In Sec.\ \ref{Sec:WKBQNM}, we calculate the QNM complex frequencies of the AOS black hole using the $6^{th}$ order Wentzel–Kramers–Brillouin (WKB) method.    In Sec.\ \ref{Sec:Analysis}, we analyze the QNM spectrum for different values of the parameter $\epsilon$ and multipole number $l$.  
In Sec.\ \ref{Sec:ringdown}, we produce and analyze the ringdown waveform for various values of $\epsilon$ and $l$.  The summary and conclusion are presented in Sec.\ \ref{Sec:conclusions}.
Finally, we include two appendices. Appendix A provides more data comparing the QNMs of different potentials and Appendix B shows the details of using the improved asymptotic iteration method and compares the results to the WKB method.

\section{Wave Equation}
\label{Sec:WE}

A massless scalar field in the background of a black hole spacetime obeys the Klein-Gordon equation
\begin{equation}
\frac{1}{\sqrt{-g}}\partial_\mu\left( {\sqrt{-g}}g^{{\mu}{\nu}}\partial_\nu{\Phi} \right)=0,
\label{KG-wave-eq}
\end{equation}
where $g_{\mu\nu}$ is the metric and $g$ is its determinant. Here we use Planck units where $c=G=\hbar=1$.  

We apply the separation of variables
\begin{equation}
	\Phi(t,r, \theta, \phi) = Y_l(\theta, \phi)\Psi(t,r)/r,
	\label{seperate-variable}
\end{equation}
where $Y_l(\theta, \phi)$ are spherical harmonics with the multipole number $l=0,1,2,\dots$, together with 
the line element of a completely general, spherically symmetric, static spacetime
\begin{equation}
ds^2=-A(r)dt^2+B(r)^{-1}dr^2+r^2d\Omega^2
\label{spherical-le}
\end{equation} 
to obtain the QNM wave equation
\beeq
\frac{\partial^2\Psi}{\partial t^2}+\left(-\frac{\partial^2}{\partial r_*^2}+V(r)\right)\Psi=0.
\label{WE-time}
\eneq  
In the above equation, $r_*$ is the tortoise coordinate linked to the radial coordinate according to
\beeq
dr_*=\frac{dr}{\sqrt{A(r)B(r)}},
\label{tortoise}
\eneq
and 
\beeq
V(r)=A(r) \frac{l(l+1)}{r^2}+ \frac{1}{2r} \frac{d}{dr} \left[ A(r)B(r) \right]
\label{scalarV}
\eneq
is the Regge-Wheeler or QNM potential.  If we assume the perturbations depend on time as 
\beeq
\Psi(t,r)=e^{-i \omega t} \psi(r),
\label{TimeDepend}
\eneq  
we obtain the time-independent wave equation
\beeq
\frac{d^2 \psi}{dr_*^2}+\left[\omega^2-V(r)\right]\psi=0,
\label{WEnoTime}
\eneq  
where $\omega$ is the complex QNM frequency to be determined after the appropriate boundary conditions are applied.

The effective field theory metric functions of the static, spherically symmetric line element (\ref{spherical-le}) for the exterior spacetime of the AOS black hole is provided explicitly by Ashtekar and Olmedo in \cite{Ashtekar3}: 
\beeq
A(r) = \left( \frac{r}{r_H} \right)^{2\epsilon} \frac{\left( 1- \left( \frac{r_H}{r} \right)^{1+\epsilon} \right) \left( 2+ \epsilon + \epsilon \left( \frac{r_H}{r} \right)^{1+\epsilon} \right)^2    \left( (2+ \epsilon)^2 - \epsilon^2 \left( \frac{r_H}{r} \right)^{1+\epsilon} \right)  
}{16 \left( 1+ \frac{\Lambda^2}{r^2} \left(\frac{r_H}{r}\right)^2\right)(1+\epsilon)^4}~
\label{Ashg00exact}
\eneq 
and
\beeq
B(r) =  \frac{\left(  \left( \frac{r}{r_H} \right)^{1+\epsilon} -1 \right) \left(  \left( \frac{r}{r_H} \right)^{1+\epsilon} (2+\epsilon)^2-\epsilon^2\right)  }{ \left( 1+ \frac{\Lambda^2}{r^2} \left(\frac{r_H}{r}\right)^2 \right) \left( \epsilon+ \left( \frac{r}{r_H} \right)^{1+\epsilon} (2+\epsilon)\right)^2}~.
\label{Ashg11exact}
\eneq 
The derivation of the AOS model provides physical interpretations for the parameters $\epsilon$ and $\Lambda$ in the quantum corrected metric, as summarized below. For more details, see \cite{Ashtekar3}.
\beeq
\epsilon \sim \frac{1}{2}\gamma^2\delta^2_b ,
\eneq
where $\gamma$ is the Barbero-Immirzi parameter determined from black hole entropy considerations to have the value $\gamma\sim 0.24$ and $\delta_b$ is determined by the black hole mass, the Barbero-Immirzi parameter and the eigenvalue $\Delta\sim 5.17 l_{\rm Pl}$ ($l_{\rm Pl}$ is the Planck length) of the fundamental area operator in LQG. Explicitly,
\beeq
\delta_b := \left( \frac{\sqrt{\Delta}}{\sqrt{2\pi}\gamma^2 m} \right)^{1/3},
\eneq
where $m$ is half the Schwarzschild radius of the black hole under consideration. The approximations used to derive the quantum corrections in the AOS metric only apply to macroscopic black holes. For $m\sim 10^6 l_{\rm Pl}$, for example, $\epsilon \sim 10^{-6}$, while for a solar mass black hole $\epsilon\sim 10^{-28}$. The effects on astrophysical observations would of course be unobservable. We take the view however that a full understanding of such quantum corrected black holes requires an investigation of features that are observable in principle if not in practice. The other quantum parameter in the corrected metric is
\beeq
\Lambda=\frac{\gamma}{8}\left(\frac{\gamma \Delta^2}{4\pi^2 m}\right)^{1/3}.
\eneq
Immediately outside the horizon of a macroscopic black hole with $r_H>10^6 l_{\rm Pl}$, say, one has
\beeq
\left(\frac{\Lambda r_H}{r^2}\right)^2 \ll  10^{-16}  \ll \epsilon.
\eneq
The term involving $\Lambda$ appears in both $A(r)$ and $B(r)$ and decreases as $1/r^4$ as one moves away from the horizon and therefore cannot affect the asymptotics and we henceforth drop it from
consideration for simplicity.


Assuming $\epsilon \ll 1$, the authors of \cite{Ashtekar3} provide the approximate metric functions 
\beeq
A(r) = \left( \frac{r}{r_H} \right)^{2\epsilon} \left( 1- \left( \frac{r_H}{r} \right)^{1+\epsilon} \right)~
\label{Ashg00approx}
\eneq 
and
\beeq
B(r) =  \left( 1- \left( \frac{r_H}{r} \right)^{1+\epsilon} \right)~
\label{Ashg11approx}
\eneq 
to probe the global structure of the spacetime outside the horizon.
In what follows, we show that the above approximation leads to a noticeably different QNM potential and, consequently, QNM frequency spectrum.  For brevity, in the remainder of this paper, we will refer to the QNM potential (\ref{scalarV})  with the metric functions  (\ref{Ashg00exact}) and (\ref{Ashg11exact}) as the AOS potential, $V_{AOS}$.  The QNM potential (\ref{scalarV})  with the approximate metric functions  (\ref{Ashg00approx}) and (\ref{Ashg11approx}) will be referred to as the approximate potential, $V_{A}$.

We point out that the Taylor series expansion of Eqs.\ (\ref{Ashg00exact}) and (\ref{Ashg00approx}) around $\epsilon=0$ only match in the $\epsilon^0$ term.  The same is true for Eqs.\ (\ref{Ashg11exact}) and (\ref{Ashg11approx}).  To resolve this discrepancy, we introduce an improved approximation where
\beeq
A(r) = \left( \frac{r}{r_H} \right)^{2\epsilon} \left( 1- \left( \frac{r_H}{r} \right)^{1+\epsilon} \right)\frac{1+\epsilon\left(1+\frac{r_H}{r} \right)}{(1+\epsilon)^3}~
\label{Raming00approx}
\eneq 
and
\beeq
B(r) =  \left( 1- \left( \frac{r_H}{r} \right)^{1+\epsilon} \right)\frac{1+\epsilon}{1+\epsilon\left(1+\frac{r_H}{r} \right)}~.
\label{Raming11approx}
\eneq 
The above functions have the same Taylor series expansion around $\epsilon=0$ as the exact expressions in (\ref{Ashg00exact}) and (\ref{Ashg11exact}) up to the order of $\epsilon$.  Note that the functions (\ref{Raming00approx}) and (\ref{Raming11approx}) are not unique and one can achieve the same Taylor series up to the order of $\epsilon$ with a variety of functions.  However,  the multiplication of the two improved approximate metric functions, $A(r)B(r)$, form a perfect square that leads to a simpler expression for the tortoise coordinate that involves $\sqrt{A(r)B(r)}$.

For the approximate metric functions (\ref{Raming00approx}) and (\ref{Raming11approx}), the QNM potential (\ref{scalarV}) takes the following form
\begin{eqnarray}
V_{IA}(r) =
\left( \frac{r}{r_H} \right)^{2\epsilon} \left( 1- \left( \frac{r_H}{r} \right)^{1+\epsilon} \right)\left(\frac{1+\epsilon\left(1+\frac{r_H}{r} \right)}{(1+\epsilon)^3}\frac{l(l+1)}{r^2} + \frac{\epsilon+ \left( \frac{r_H}{r} \right)^{1+\epsilon}}{r^2(1+\epsilon)^2} \right)~,
\label{Vqnm}
\end{eqnarray}
where $IA$ in the subscript stands for improved approximation.  For $\epsilon=0$, the above potential reduces to the Schwarzschild QNM potential for scalar perturbations. Using the approximation $V_{IA}$, in place of the exact potential $V_{AOS}$, significantly simplifies both analytical and numerical calculations.  We find the speed-up in numerical calculations indispensable, for example calculation times of days versus weeks.

\begin{figure}[th!]
	\begin{center}
		\includegraphics[height=5.1cm]{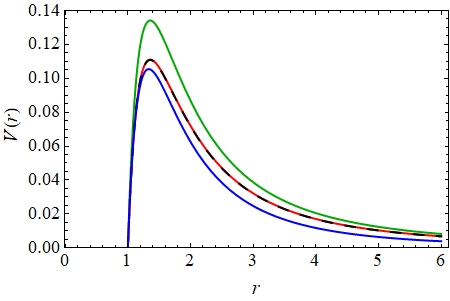}
		\hspace{0.3cm}
		\includegraphics[height=4.95cm]{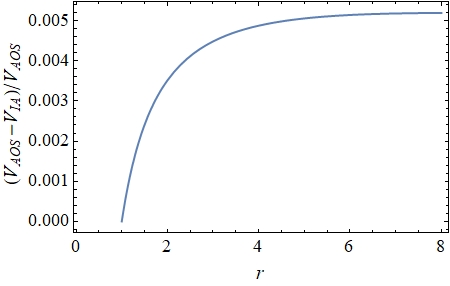}
	\end{center}
	\vspace{-0.7cm}
	\caption{\footnotesize The left plot shows the QNM potential versus radial coordinate for $l=0$, $\epsilon=0.1$ and $\Lambda=0$.  For comparison, we include $V_{AOS}$ in dashed black, the tallest potential $V_A$ in green, and $V_{IA}$ in red.  We also include the Schwarzschild QNM potential ($\epsilon=\Lambda=0$), which is the shortest, in blue.  We use units where $r_H=1$. $V_{AOS}$ and $V_{IA}$ show almost perfect agreement. The plot to the right shows that the relative error of $V_{IA}$ compared to $V_{AOS}$ is small ($\lesssim 0.005$).}
	\label{V(r)}
\end{figure}

To compare, in 
Figure \ref{V(r)}, we plot all the three QNM potentials, i.e.\ $V_{AOS}$, $V_A$ and $V_{IA}$, together with the Schwarzschild QNM potential.  It is noteworthy that despite the divergence of the metric function $A(r)$ as $r \rightarrow \infty$, which has been argued to be unsettling in \cite{Bouhmadi, Faraoni}, the QNM potential of this black hole model is well behaved as $r \rightarrow \infty$ as can be seen in Figure \ref{V(r)}.  The two QNM potentials $V_{AOS}$ and $V_{IA}$ are almost identical, but $V_A$ is significantly different than all the other cases in Figure \ref{V(r)}.
Since $V_A$ is very different
from the other two potentials outside of the horizon, it is not suitable for QNM calculations. Comparison of
the complex QNM frequencies of the three potentials in Appendix A confirms this.
Note however that while $V_A$ cannot be used in QNM calculations, the metric functions (\ref{Ashg00approx}) and (\ref{Ashg11approx}) have the right asymptotic behavior to be used in the calculations in \cite{Ashtekar3}. 

The tortoise coordinate for this spacetime can be derived by combining Eqs.~(\ref{tortoise}), (\ref{Raming00approx}) and (\ref{Raming11approx}):
\begin{eqnarray}
r_* = \int \frac{(1+\epsilon)dr}{\left( \frac{r}{r_H} \right)^{\epsilon} \left( 1- \left( \frac{r_H}{r} \right)^{1+\epsilon} \right)}=-\frac{(1+\epsilon)}{2}\frac{r^2}{r_H} ~_2F_1\left( 1, \frac{2}{1+\epsilon}; 1+\frac{2}{1+\epsilon}; \left(\frac{r}{r_H}\right)^{1+\epsilon} \right) + C ~,
\label{r-tortoise}
\end{eqnarray}
where  $_2F_1(a,b;c;z)$ is the hypergeometric function and $C$ is the constant of integration, which should be chosen such that the tortoise coordinate has no imaginary component.  In the rest of this paper, we also choose $C$ so that the peak of the QNM potential in the tortoise coordinate is centered at $r_*=0$.  Note that when $\epsilon=0$, the relationship between the tortoise coordinate and the radial coordinate reduces to the Schwarzschild case where $r_*=r+r_H \ln(r-r_H)+C$.

With our choice of time-dependence (\ref{TimeDepend}), the boundary conditions at the event horizon and infinity are, respectively,
\beeq
\begin{array}{ll}
	\psi(x) \rightarrow e^{-i\omega r_*}  & \mbox{as $r_* \rightarrow -\infty$ ($r \rightarrow r_H$)}~,\\
	\psi(x) \rightarrow e^{i\omega r_*}  & \mbox{as $r_* \rightarrow \infty$ ($r \rightarrow \infty$)}~.
\end{array}       
\label{B.C.}
\eneq
For the purpose of brevity, in the rest of this paper, we choose units in which $r_H=1$.  That means $r$, $r_*$ and $t$ are expressed in units of $r_H$. The QNM frequency $\omega$ is in units of $r_H^{-1}$ and the units for the QNM potential $V(r)$ are $r_H^{-2}$.

\section{The WKB Method}
\label{Sec:WKBQNM}

To calculate the QNMs of the AOS black hole, we use the WKB method.  This method was originally applied to the problem of black hole perturbations by Schutz and Will in \cite{WKB1}.  The formula for the $3^{rd}$ order WKB method is derived in \cite{WKB2} and it is extended to the $6^{th}$ order by Konoplya in \cite{WKB-Konoplya}.  

To determine the QNM spectrum using the WKB method, one needs to solve the following equation\cite{WKB2}
\beeq
\frac{i \left[ \omega^2 -V(r_{*})|_{\bar{r}_*}\right]}{\sqrt{2V''(r_*)|_{\bar{r}_*}}} -\overset{N}{\underset{j=2}{\sum}} \Lambda_j(n)=n+\dfrac{1}{2},
\label{WKBorder}
\eneq
where $\bar{r}_*$ is the location of the maximum of the QNM potential $V(r_*)$ in the tortoise coordinate. Prime indicates differentiation with respect to $r_*$,  and $\Lambda_j(n)$ are the WKB correction terms. $N$ indicates the order of the WKB method.  $\Lambda_{2,3}$ are given in \cite{WKB2}\footnote{$\Lambda_2$ in \cite{WKB2} is missing a factor of $i$ in the numerator.} and $\Lambda_{4,5,6}$ can be found in \cite{WKB-Konoplya}.

We find that for $\epsilon=0.1$, the QNMs of the potential with the improved approximation, $V_{IA}$, match those of the QNMs of the AOS potential, $V_{AOS}$, up to three significant figures.  The accuracy further improves for $\epsilon=0.01$ where the QNMs of the two potentials ($V_{IA}$ and $V_{AOS}$) match up to five significant figures. This shows that the improved approximation produces accurate results for small $\epsilon$.  
Tables comparing QNMs of these potentials appear in Appendix A.

The AOS metric functions (\ref{Ashg00exact}) and (\ref{Ashg11exact}) are far more complicated than the metric functions (\ref{Raming00approx}) and (\ref{Raming11approx}) in the improved approximation.  The calculations are analytically more complex and numerically more intense for the AOS metric functions.  Since the difference in the results is negligible for $\epsilon \lesssim 0.1$, as is evident in the results presented in Appendix A, it makes more sense to use the improved approximation.  Thus, in the rest of this paper, we focus on the improved approximation.

In Table I, we provide the QNM complex frequencies for different values of the overtone number $n$ and the multipole number $l$ for the potential $V_{IA}$.  As indicated in \cite{WKB-Konoplya}, the WKB method works more accurately for lower values of $n$ and higher values of $l$.   For example, for $l=0,1,2$ we only find less than six reliable roots, while for $l=6$ we find twelve reliable roots.

\vspace{0.3cm}
\footnotesize
\begin{tabular}{cccccc}
	\multicolumn{6}{c}{Table I:  $\omega$ for $V_{IA}$ for different values of $\epsilon$ using $6^{th}$ order WKB method\cite{KonoplyaCode}} \\ 
	\hline
	$n,l$ & $\epsilon=0$ &  $\epsilon=0.01$ & $\epsilon=0.1$    &\\ 
	\hline 
	0,0 & $0.22093 - 0.20164i$ &  $0.22227 - 0.20108 i$ & $ 0.23481 - 0.19592 i$  \\ 
	1,0 & $0.17805 - 0.68910i$ &  $0.17860 - 0.68812i$ & $0.18300 - 0.67919i$  \\ 
	\hline 
	0,1 & $0.58582 - 0.19552i$ &  $0.58703 - 0.19535i$ & $0.59765 - 0.19339 i$  \\ 
	1,1 & $0.52894 - 0.61304i$ &  $0.52996 - 0.61249i$ & $0.53857 - 0.60636i$   \\ 
	2,1 & $0.46203 - 1.0843i$ &  $0.46243 - 1.0837i$ & $0.46446 - 1.0764 i$  \\
	\hline 
	0,2 & $0.96728 - 0.19353i$ &  $0.96890 - 0.19337i$ & $0.98272 - 0.19152 i$  \\ 
	1,2 & $0.92769 - 0.59125i$ &  $0.92921 - 0.59075i$ & $0.94205 - 0.58501 i$   \\ 
	2,2 & $0.86077 - 1.0174i$ &  $0.86202 - 1.0166i$ & $0.87196 - 1.0071i$ \\
	3,2 & $0.78641 - 1.4798i$ &  $0.78712 - 1.4788 i$ & $0.79126 - 1.4673i$ \\
	4,2 & $0.72517 - 1.9766i$ &  $0.72496 - 1.9759 i$ & $0.71945 - 1.9670 i$  \\
    \hline 
	0,6 & $2.5038 - 0.19261i$ &  $2.5074 - 0.19246 i$ & $2.5383 - 0.19067  i$  \\ 
	1,6 & $2.4875 - 0.57947i$ &  $2.4911 - 0.57900i$ & $2.5217 - 0.57360i$   \\ 
	2,6 & $2.4557 - 0.97120i$ &  $2.4593 - 0.97041 i$ & $2.4891 - 0.96129  i$  \\
	3,6 & $2.4098 - 1.3708i$ &  $2.4132 - 1.3697 i$ & $2.4418 - 1.3568i$  \\
	4,6 & $2.3519 - 1.7809i$ &  $2.3552 - 1.7795  i$ & $2.3818 - 1.7627 i$  \\
	5,6 & $2.2848 - 2.2036i$ &  $2.2878 - 2.2018i$ & $2.3117 - 2.1814i$  \\
	6,6 & $2.2113 - 2.6402i$ &  $2.2140 - 2.6381i$ & $2.2341 - 2.6144 i$  \\
	7,6 & $2.1347 - 3.0914i$ &  $2.1369 - 3.0891 i$ & $2.1521 - 3.0628 i$  \\
	8,6 & $2.0579 - 3.5575i$ &  $2.0595 - 3.5551 i$ & $2.0686 - 3.5271 i$  \\
	9,6 & $1.9839 - 4.0383 i$ &  $1.9848 - 4.0359 i$ & $1.9863 - 4.0077  i$  \\
	10,6 & $1.9155 - 4.5334 i$ &  $1.9155 - 4.5312 i$ & $1.9080 - 4.5047  i$  \\
	11,6 & $1.8555 - 5.0421i$ &  $1.8546 - 5.0403 i$ & $1.8363 - 5.0180 i$  \\
	\label{Table1}
\end{tabular} 
\normalsize
\vspace{-0.cm}

\begin{figure}[th!]
	\begin{center}
		\includegraphics[height=3.9cm]{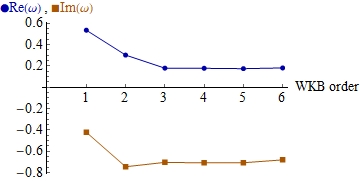}
		\includegraphics[height=3.9cm]{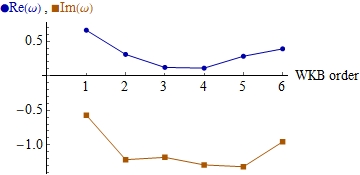}
		\includegraphics[height=3.9cm]{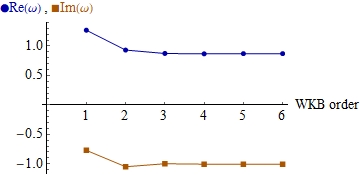}
	\end{center}
	\vspace{-0.7cm}
	\caption{\footnotesize The real (top in blue) and imaginary (bottom in brown) parts of $\omega$ as a function of the order of the WKB method for $\epsilon=0.1$.  On the top left $l = 0$ and $n = 1$. On the top right $l=0$ and $n=2$. On the lower plot $l=2$ and $n=2$.\cite{KonoplyaCode}}
	\label{w-WKBorder}
\end{figure}
\vspace{-0.2cm}

The reliability of the roots is determined by comparing the results for different orders of the WKB method.   For example, in Figure \ref{w-WKBorder} we see the real and imaginary parts of $\omega$ converge as we increase the order of the WKB method for $l=0$ and $n=1$.  That is not the case for  $l=0$ and $n=2$. Convergence for higher values of $n$ improves with higher values of $l$.  We illustrate this in Figure \ref{w-WKBorder}, where we plot the real and imaginary parts of $\omega$ for $l = 2$ and $n = 2$ to be compared with the case of $l=0$ and $n=2$.

We also look at convergence plots, similar to Figure \ref{w-WKBorder}, to determine if the accuracy of the WKB method depends on the value of the parameter $\epsilon$.  We find no correlation.

As mentioned earlier, the WKB method becomes less accurate for lower values of $l$
and higher values of $n$. Since the accuracy of the WKB method is in question, it is
useful to compare the results to those found using other methods. We have therefore used
the improved asymptotic iteration method (AIM) to determine QNM frequencies for the
potential $V_{IA}$ provided in Eq.\ (\ref{Vqnm}). The results are in good agreement with
the WKB method. In Appendix B we describe the AIM briefly and show how to
adapt the AIM to our black hole model. We also present the numerical results to show the
consistency of the QNMs obtained using two different numerical methods.



\section{An Analysis of the QNM Spectrum}
\label{Sec:Analysis}

\begin{figure}[th!]
	\begin{center}
		\includegraphics[height=8.cm]{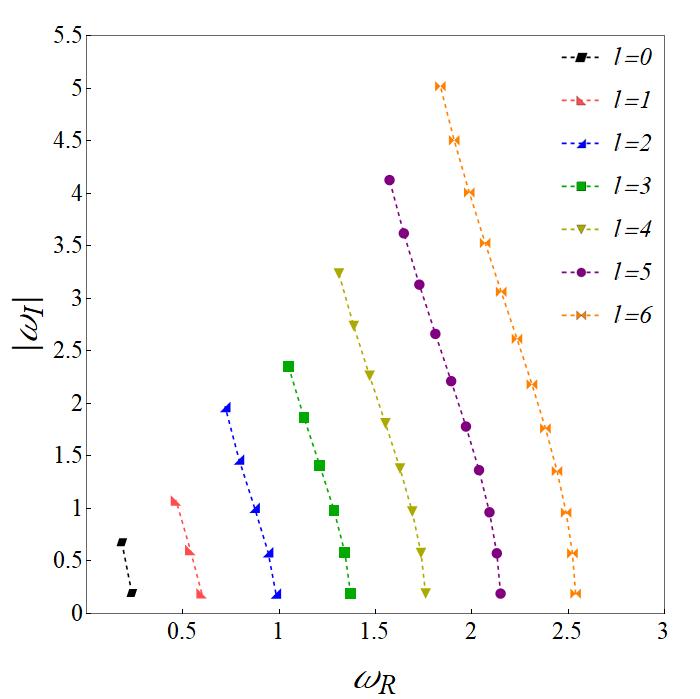}
	\end{center}
	\vspace{-0.7cm}
	\caption{\footnotesize Scalar QNM spectrum for $\epsilon=0.1$ for different values of $l$.  The roots are generated for $V_{IA}$ using the WKB method.}
	\label{WKB-different-l}
\end{figure}

In Figure \ref{WKB-different-l}, we plot the QNM spectrum for $\epsilon=0.1$ for different values of $l$.  As $l$ increases, the real part of the $n^{\text{th}}$ QNM frequency ($\omega_R$) increases, by a nearly fixed amount, and there is a small decrease in the magnitude of the imaginary part ($|\omega_I|$). 

\begin{figure}[th!]
	\begin{center}
		\includegraphics[height=8.cm]{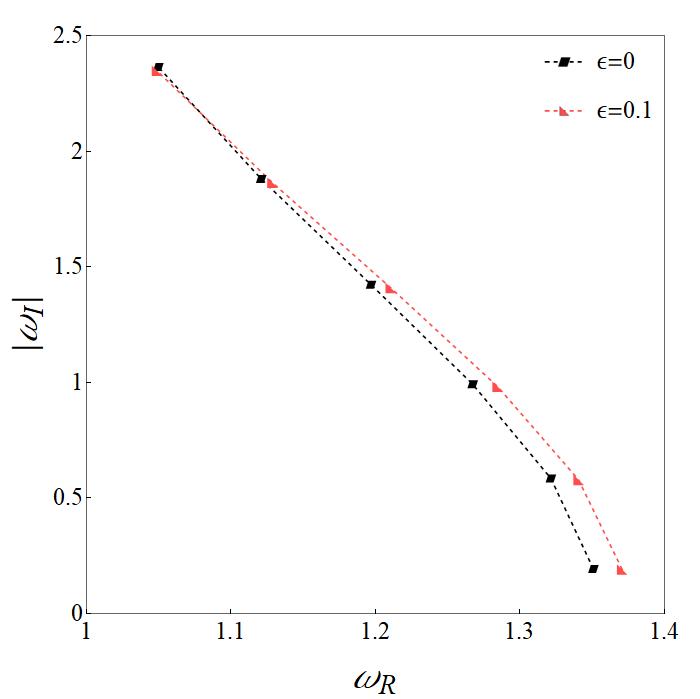}
		\includegraphics[height=8.cm]{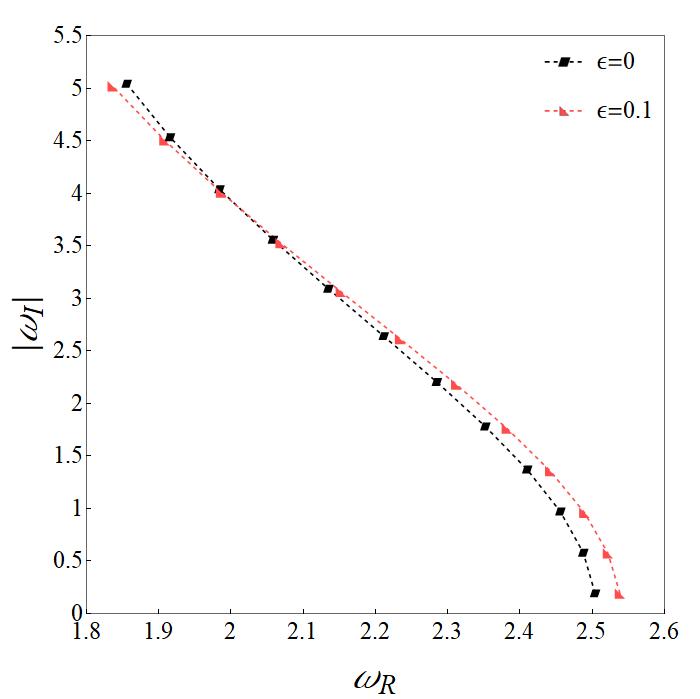}
	\end{center}
	\vspace{-0.7cm}
	\caption{\footnotesize Scalar QNM spectrum for $l=3$, on the left, and  $l=6$, on the right. In both graphs we include the cases $\epsilon=0$ (Schwarzschild) and $\epsilon=0.1$ for $V_{IA}$. The roots are generated using the WKB method. }
	\label{different-e}
\end{figure}

In Figure \ref{different-e}, we compare the QNM spectrum of the AOS black hole for $\epsilon=0.1$ with the Schwarzschild spectrum ($\epsilon=0$) for $l=3$ and $l=6$.  In both graphs, we include the roots generated by the WKB method.  In cases where we were able to use the AIM, the roots from both methods match well. For $\epsilon=0.1$, the real part of the low damping QNMs is larger than the Schwarzschild modes, but $\omega_R$ becomes smaller than the Schwarzschild case for higher overtone QNMs.  This is different than the behavior of the PK black hole QNM spectrum \cite{DGMK}, which stays parallel to the Schwarzschild spectrum.  The damping rate ($|\omega_I|$) of the AOS black hole QNM spectrum is slightly lower than the Schwarzschild spectrum.

\newpage
\section{The Ringdown Waveform}
\label{Sec:ringdown}

To generate the ringdown waveform, we numerically solve the time-dependent wave equation (\ref{WE-time}) using the initial data
\beeq
\Psi(r_*,0)={\cal A} \exp \left(- \frac{(r_*-\bar{r}_{*})^2}{2\sigma^2} \right),~  \partial_t \Psi|_{t=0}=-\partial_{r_*} \Psi(r_*, 0)~,
\label{GaussianWave}
\eneq  
where we use $\sigma=1$, $\bar{r}_*=-40$, and ${\cal A}=20$.  We choose the observer to be located at $r_*=90$.   To carry out the calculations, we use the built-in {\em Mathematica} commands for solving partial differential equations.

\begin{figure}[th!]
	\begin{center}
		\includegraphics[height=5.3cm]{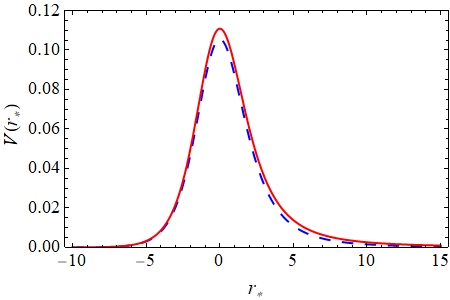}
		\hspace{0.1cm}
		\includegraphics[height=5.15cm]{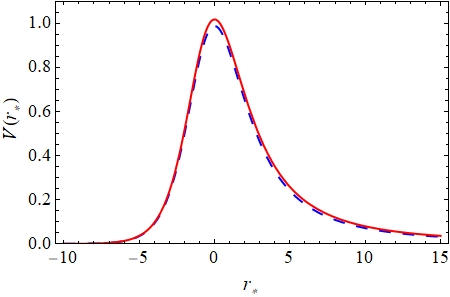}
	\end{center}
	\vspace{-0.7cm}
	\caption{\footnotesize QNM potential versus tortoise coordinate for $l=0$ (left) and $l=2$ (right).  For comparison, we include the Schwarzschild ($\epsilon=\Lambda=0$) potential in dashed blue and $V_{IA}$ for $\epsilon=0.1$ in solid red.}
	\label{V(rstar)}
\end{figure}

In Figure \ref{V(rstar)}, we compare the shape of the Schwarzschild potential ($\epsilon=\Lambda=0$) in the  tortoise coordinate for $l=0$ and $l=2$ with the shape of $V_{IA}$ for $\epsilon=0.1$.  We do not plot $V_{AOS}$ since it is indistinguishable from $V_{IA}$. 

\begin{figure}[th!]
	\begin{center}
		\includegraphics[height=5.2cm]{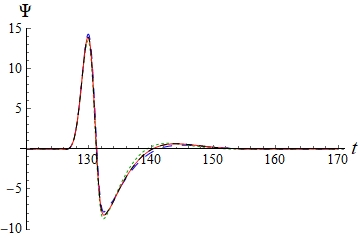}
		\includegraphics[height=5.2cm]{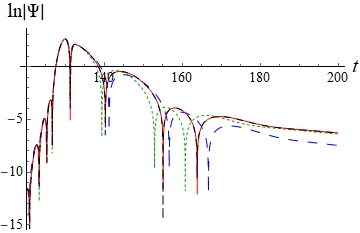}
	\end{center}
	\vspace{-0.7cm}
	\caption{\footnotesize $\Psi$ (left) and $\ln |\Psi|$ (right) as a function of time for $l=0$.  In both graphs, we include the cases $\epsilon=0$ (Schwarzschild) in dashed blue and $\epsilon=0.1$ in dashed black for $V_{AOS}$, solid red for $V_{IA}$, and dotted green for $V_{A}$. $\Lambda$ is taken to be zero. The ringdown waveform for $V_{AOS}$ and $V_{IA}$ show almost perfect agreement.}
	\label{wave-l0}
\end{figure}

\begin{figure}[th!]
	\begin{center}
		\includegraphics[height=5.2cm]{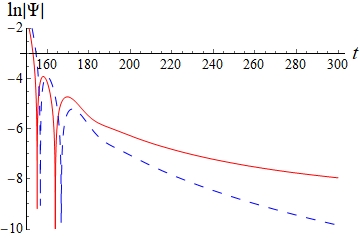}
		\hspace{0.2cm}
		\includegraphics[height=5.2cm]{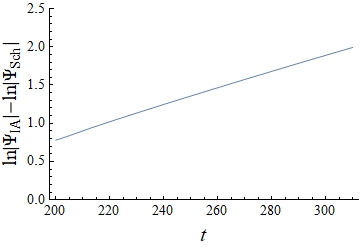}
	\end{center}
	\vspace{-0.7cm}
	\caption{\footnotesize On the left, we show more of the tail behavior of $\ln|\Psi|$ with $l=0$ for  $\epsilon=0$ (Schwarzschild), in dashed blue, and $\epsilon=0.1$, in solid red, for $V_{IA}$. The tails are diverging.  To make this clear, on the right, we show the difference in $\ln|\Psi|$ between Schwarzschild and $V_{IA}$ for large $t$.}
	\label{l=0-Slope}
\end{figure}

In Figure \ref{wave-l0} we plot the ringdown waveform $\Psi$ and $\ln|\Psi|$, as a function of time, for $l=0$ for all the three potentials $V_{AOS}$, $V_{IA}$, and $V_A$.  We also include the Schwarzschild ringdown waveform for comparison.  $V_{AOS}$ and $V_{IA}$ produce almost identical waveforms, but the waveform produced by $V_A$ is visibly different.  The oscillation periods are easier to see in the log plot, where it is clear that the oscillation frequency is the lowest for the Schwarzschild case and highest for $V_A$. Another interesting feature observed in Figure \ref{wave-l0} is the behavior of the power-law tail of the ringdown waveform, where the Schwarzschild case is different from all the other cases. Figure \ref{l=0-Slope} shows more of the tail where it is easier to see that the tails are diverging.  To make this clear, in Figure \ref{l=0-Slope}, we also plot the difference in $\ln|\Psi|$ between Schwarzschild and $V_{IA}$ for large $t$, which has a linear behavior.  This indicates that the difference in the slope of the tails of Schwarzschild and $V_{IA}$ stays constant for large $t$. This does not occur when $l\neq0$ or for other RBHs, for which the waveform asymptotes to Schwarzschild. Studies of the Schwarzschild spacetime show a link between the power-law tail and the scattering waves at large radius\cite{Price}.  Therefore, it is reasonable to infer that the difference in the power-law tail is due to the asymptotic behavior of the QNM potential of the Schwarzschild black hole, which drops as $1/r^3$ for $l=0$, while $V_{AOS}$, $V_{IA}$ and $V_A$ all drop as $\approx 1/r^2$.  


\begin{figure}[th!]
	\begin{center}
		\includegraphics[height=5.2cm]{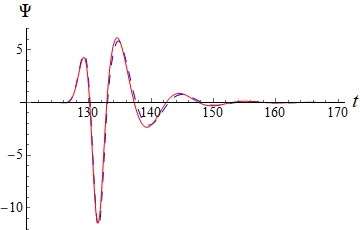}
		\includegraphics[height=5.2cm]{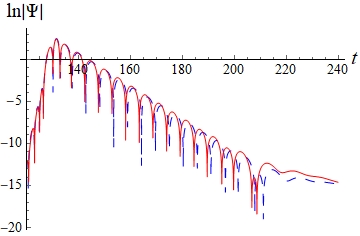}
	\end{center}
	\vspace{-0.7cm}
	\caption{\footnotesize $\Psi$ (left) and $\ln |\Psi|$ (right) as a function of time for $l=1$.  In both graphs, we include the cases $\epsilon=0$ (Schwarzschild) in dashed blue and $\epsilon=0.1$  in solid red for $V_{IA}$.}
	\label{wave-l1}
\end{figure}

\begin{figure}[th!]
	\begin{center}
		\includegraphics[height=5.2cm]{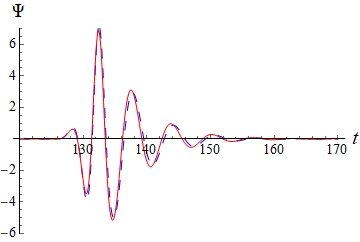}
		\includegraphics[height=5.2cm]{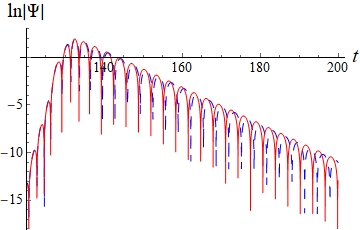}
	\end{center}
	\vspace{-0.7cm}
	\caption{\footnotesize $\Psi$ (left) and $\ln |\Psi|$ (right) as a function of time for $l=2$.  In both graphs, we include the cases $\epsilon=0$ (Schwarzschild) in dashed blue and $\epsilon=0.1$  in solid red for $V_{IA}$.}
	\label{wave-l2}
\end{figure}

In Figures \ref{wave-l1} and \ref{wave-l2} we plot the ringdown waveform $\Psi$ and $\ln|\Psi|$, as a function of time, for $l=1$ and $l=2$ for the Schwarzschild potential and $V_{IA}$. In the log plot, it is clear that the oscillation frequency is higher for the AOS black hole.


To further check for the consistency between the numerically generated ringdown waveforms and QNM data provided in Tables III and IV, we use the Prony method\cite{Prony} to extract the first ($n=0$) QNM frequency from the waveforms generated for $\epsilon=0.1$ shown in Figures \ref{wave-l0}, \ref{wave-l1} and \ref{wave-l2}.  
In the case of $l=0$, we find $0.23334-0.20535i$ for $V_{IA}$ and $0.23373-0.20545i$ for $V_{AOS}$. For $l=1$ and $l=2$, the results are $0.59776-0.19300i$ and $0.98265-0.19145i$ respectively for the potential $V_{IA}$.  These are all in good agreement with the data presented in Tables III and IV.

\section{Summary and Conclusion}
\label{Sec:conclusions}

We studied massless scalar field perturbations in the background of the exterior of the AOS black hole.  The spacetime of this black hole is rather unique in the sense that, in Schwarzschild coordinates, one of the metric functions diverges as $r\to \infty$ even though it was shown in \cite{Ashtekar3} that the spacetime is asymptotically flat. We showed that despite this unusual asymptotic behavior, the QNM potential $V_{AOS}$ is well defined everywhere when Schwarzschild coordinates are used.  

In addition to calculating the QNM spectrum and the ringdown waveform of the effective field theory metric, we found a useful new approximate form of the metric.  This new approximation makes the calculations significantly easier and is  more suitable for such calculations than the previous approximation used in \cite{Ashtekar3} to probe the global structure of the spacetime.  While the calculations are, in principle, possible using the exact form of the metric,  our approximation allows us to produce QNM frequencies and ringdown waveforms to high accuracy with manageable computation times. In the case of AIM,  our approximation ($V_{IA}$) led to highly complicated expressions.  The AIM may be intractable using the exact potential $V_{AOS}$.

We calculated the QNMs of the AOS black hole using the $6^{th}$ order WKB method and the AIM.  Both methods gave consistent results.  The $6^{th}$ order WKB method was applied to $V_{AOS}$, $V_A$, and $V_{IA}$.  The QNM spectra of $V_{AOS}$ and $V_{IA}$ are almost identical for $\epsilon \lesssim 0.1$ and they both differ form the QNM spectrum produced by $V_{A}$. The AIM was applied to $V_{IA}$ and consistent results were found.   We also examined the ringdown waveform of this black hole for all three cases ($V_{AOS}$, $V_A$ and $V_{IA}$) and compared all our results with the Schwarzschild case.  Once again, the ringdown waveform produced by $V_{AOS}$ and $V_{IA}$ are almost identical and they differ from $V_A$ and the Schwarzschild case.  As a consistency check, we calculated the least damped QNM ($n=0$) of the ringdown waveforms using the Prony method and we found consistent results.



The QNM frequencies of the AOS black hole follow closely the QNM spectrum of a Schwarzschild black hole.  We found no modes with positive damping, which indicates such a RBH is stable against massless scalar perturbations.  
We showed that an increase in the magnitude of $\epsilon$ increases the height of the QNM potential and gives oscillations with higher frequency and less damping.

An interesting aspect of the AOS black hole is the asymptotic behavior of corrections to the QNM potential, which drop off as $\approx 1/r^2$. 
When $l=0$, the Schwarzschild QNM potential drops as $1/r^3$, so the $1/r^2$ correction dominates the potential for large $r$.  For non-zero $l$, the correction is of the same order as the leading $r \rightarrow \infty$ term.  This is in contrast to other regular asymptotically Schwarzschild black holes for which the quantum correction becomes negligible compared to the classical terms at large $r$. It seems reasonable to draw the conclusion that this $1/r^2$ correction is responsible for the qualitative difference observed in the power-law tail in Figure  \ref{wave-l0} between AOS and Schwarzschild waveforms at large times.  This conclusion is supported by the studies of the Schwarzschild spacetime, which show that the power-law tail is caused by the scattering waves off the potential at large radius\cite{Price}.  This suggests the quantum corrections affect the behavior of the waves for large $r$.



\vskip .5cm

\leftline{\bf Acknowledgments}
We thank Wei-Liang Qian for sharing with us the code for the Prony method.  The work of G.K.\ was funded in part by the Natural Sciences and Engineering Research Council of Canada.

\appendix
\section{Numerical Data for Different Potentials}
In Tables II and III, for comparison, we provide the QNM complex frequencies for the multipole numbers $l=0$  and $l=3$ for the three different potentials $V_{AOS}$, $V_A$ and $V_{IA}$.  The value of $\epsilon$ in Table I is $0.1$ and in Table II is $0.01$.  Note the results for $V_{IA}$ match well with $V_{AOS}$.  However, the approximate potential $V_A$ matches only up to one significant figure for $\epsilon=0.1$ and only up to two significant figures for $\epsilon=0.01$.

\vspace{0.4cm}
\footnotesize
\begin{tabular}{cccccc}
	\multicolumn{6}{c}{Table II:  $\omega$ for $\epsilon=0.1$ for the three different potentials using $6^{th}$ order WKB method\cite{KonoplyaCode}} \\ 
	\hline
	$n,l$ & $V_{AOS}$ & $V_{IA}$ &  $V_A$  &\\ 
	\hline 
	0,0 &  $ 0.235193 - 0.196029 i$  &   $0.234806 - 0.195919i$ &  $0.258287 - 0.215511 i$\\ 
	1,0 & $0.183430 - 0.679378i$    &  $0.182999 - 0.679192$ &  $0.201299 - 0.747111i$\\ 
	\hline 
	0,3 &  $1.37350 - 0.191270 i$  &  $1.37062 - 0.191030i$ &  $1.46611 - 0.208058 i$\\ 
	1,3 & $1.34342 - 0.579258i$  &     $1.34052 - 0.578534i$ &  $1.43196 - 0.630346i$\\ 
	2,3 & $1.28746 - 0.982962 i$   & $1.28450 - 0.981761i$  &  $1.36835 - 1.07050 i$\\
	3,3 & $1.21360 - 1.41032 i$   &  $1.21057 - 1.40866 i$ &  $1.28428 - 1.53766  i$\\
	4,3 & $1.13183 - 1.86567 i$   &  $1.12868 - 1.86357 i$ &  $1.19091 - 2.03699  i$\\
	5,3 & $1.05210 - 2.35036 i$   & $1.04880 - 2.34787i$  &  $1.09928 - 2.57050i$\\
	\label{Table1}
\end{tabular} 
\normalsize

\vspace{-0.cm}
\footnotesize
\begin{tabular}{cccccc}
	\multicolumn{6}{c}{Table III:  $\omega$ for $\epsilon=0.01$ for the three different potentials using  $6^{th}$ order WKB method\cite{KonoplyaCode}} \\ 
	\hline
	$n,l$ & $V_{AOS}$  & $V_{IA}$  &  $V_A$ &\\ 
	\hline 
	0,0 & $ 0.222271 - 0.201083 i$  &   $0.222265 - 0.201083 i$ &  $0.224488 - 0.203094  i$ \\ 
	1,0 & $0.178608 - 0.688112i$   &   $0.178600 - 0.688124i$ &  $0.180386 - 0.695005i$\\ 
	\hline 
	0,3 & $1.35286 - 0.192846 i$  &  $1.35283 - 0.192843i$  &  $1.36199 - 0.194566 i$\\ 
	1,3 & $1.32341 - 0.584098 i$   &   $1.32338 - 0.58409 i$ &  $1.33214 - 0.589335i$\\ 
	2,3 & $1.26910 - 0.991208 i$   &   $1.26906 - 0.991195i$ &  $1.27708 - 1.00019 i$\\
	3,3 & $1.19847 - 1.42165 i$   &  $1.19843 - 1.42163 i$ &  $1.20547 - 1.43471  i$\\
	4,3 & $1.12197 - 1.87886 i$   &  $1.12193 - 1.87883 i$  &  $1.12790 - 1.89639  i$\\
	5,3 & $1.05000 - 2.36289 i$   & $1.04996 - 2.36286 i$  &  $1.05490 - 2.38535i$\\
	\label{Table1}
\end{tabular} 
\normalsize
\vspace{-0.2cm}

\section{The Improved Asymptotic Iteration Method}
\label{Sec:IAIM}


In this section, we describe the improved asymptotic iteration method (AIM) \cite{AIM} and use it to determine QNM frequencies for the potential $V_{IA}$ provided in Eq.\ (\ref{Vqnm}).  

The AIM is useful for studying linear second order differential equations of the form
\begin{equation}
	\chi'' = \lambda_0(x) \chi' + s_0(x) \chi .
	\label{AIMequation}
\end{equation}
For such equations, the higher derivatives of $\chi$ can be expressed in terms of $\chi'$ and $\chi$ as
\begin{equation}
	\chi^{(n+1)} = \lambda_{n-1}(x) \chi' + s_{n-1}(x) \chi ,
\end{equation}
where
\begin{gather}
	\lambda_n(x) = \lambda_{n-1}'(x) + s_{n-1}(x) + \lambda_0(x)\lambda_{n-1}(x) \nonumber \\
	s_n(x) = s_{n-1}'(x) + s_0(x)\lambda_{n-1}(x). 
\end{gather}
One can then expand $\lambda_n$ and $s_n$ in a Taylor series around some point $x_0$:
\begin{gather}
	\lambda_n(x) = \sum_{i=0}^{\infty} c_n^i(x-x_0)^i \nonumber\\
	s_n(x) = \sum_{i=0}^{\infty} d_n^i(x-x_0)^i.
	\label{Taylor-ls}
\end{gather}
The recurrence relations for $\lambda_n$ and $s_n$ can now be written in terms of the coefficients $c_n$ and $d_n$ as follows:
\begin{gather}
	c_n^i = (i+1)c_{n-1}^{i+1} + d_{n-1}^i + \sum_{k=0}^i c_0^k c_{n-1}^{i-k}  \\
	d_n^i = (i+1)c_{n-1}^{i+1} + \sum_{k=0}^i d_0^k c_{n-1}^{i-k} .
\end{gather}
We then make the assumption that, for large $n$,
\begin{equation}
	\frac{s_n(x)}{\lambda_n(x)} = \frac{s_{n-1}(x)}{\lambda_{n-1}(x)}. 
	\label{AIMeq}
\end{equation}
After combining Eqs.\ (\ref{Taylor-ls})-(\ref{AIMeq}), one obtains an equation in terms of the Taylor series coefficients:
\begin{equation}
	d_n^0 c_{n-1}^0 - d_{n-1}^0 c_n^0 = 0.
	\label{recurrenceforcoeffs}
\end{equation}
The QNM frequency spectrum can be determined by solving the above equation.

To implement the AIM, first we find the asymptotic behavior of the solution to the wave equation (\ref{WEnoTime}) for $V(r)=V_{IA}(r)$ at the boundaries:
\beeq
\psi \overset{r \rightarrow 1}{\larrow}  \left(r^{1+\epsilon}-1 \right)^{-i \omega}~~\mbox{and}~~\psi \overset{r \rightarrow \infty}{\larrow} r^{i\omega(1-\epsilon)} e^{i\omega  \frac{1+\epsilon}{1-\epsilon}r^{1-\epsilon}}~.
\label{}
\eneq
We then scale out the asymptotic behavior using
\beeq
\psi(r) =  \left( \frac{r^{1+\epsilon}-1}{r^{1+\epsilon}} \right)^{-i \omega}  
r^{i\omega(1-\epsilon)} e^{i\omega  \frac{1+\epsilon}{1-\epsilon}\left( \frac{r-1}{r^\epsilon} \right)} \chi(r).
\label{WaveF-X}
\eneq
The AIM requires a finite domain so we introduce the change of variable\cite{AIM}
\begin{equation}
	\xi = 1- \frac{1}{r^{1+\epsilon}}~,
	\label{xi}
\end{equation}
which transforms the domain $\left[1,\infty \right)$ to $\left[0,1\right)$.
Combining Eqs.~(\ref{WEnoTime}), (\ref{WaveF-X}) and (\ref{xi}) leads to a differential equation for $\chi(\xi)$ in the form of Eq.\ (\ref{AIMequation}) where 
\begin{equation}
	\lambda_0 = -\frac{1}{(1-\xi)} \left( \frac{2i\omega}{ (1-\xi)^{\frac{1-\epsilon}{1+\epsilon}}}  +2i\omega \frac{\epsilon(1-\xi)^{\frac{\epsilon}{1+\epsilon}}}{1-\epsilon} + \frac{1-2i\omega}{\xi}  - \frac{3+\epsilon-4i\omega}{1+\epsilon}\right)
\end{equation}
and
\begin{eqnarray}
	s_0 &=& \frac{(1+\epsilon)^2-\xi(1+\epsilon)+\left(1+\epsilon+\epsilon(1-\xi)^{\frac{1}{1+\epsilon}}\right)l(l+1)}
	{\xi(1-\xi)^2(1+\epsilon)^3}
	-\frac{\omega^2}{\xi^2(1-\xi)^{\frac{4}{1+\epsilon}}}  \nonumber \\
	&& + 	\frac{i \omega (1-\xi)^{\frac{2\epsilon}{1+\epsilon}}}{\xi^2(1-\epsilon^2)^2} \left\{ \frac{4(1-\epsilon)^2(1-i \omega)}{ (1-\xi)^{\frac{2\epsilon}{1+\epsilon}}} + \frac{(1-\epsilon)^4(1-i \omega)}{ (1-\xi)^{4-\frac{2}{1+\epsilon}}}  + \frac{\epsilon(1-\epsilon^2)(2+\epsilon-4i\omega)}{ (1-\xi)^{\frac{\epsilon}{1+\epsilon}}} \right. \nonumber \\
	&&  \left.  -\frac{(1-\epsilon)^2[5+\epsilon(-2+\epsilon+4i\omega)-4i\omega]}{ (1-\xi)^{3-\frac{2}{1+\epsilon}}} 
	- \frac{2i\omega(1-\epsilon)^3(1+\epsilon)}{ (1-\xi)^3}  
	-\frac{2i\omega \epsilon (1-\epsilon)(1+\epsilon)^2}{ (1-\xi)^{\frac{1}{1+\epsilon}}} \right.  \nonumber \\
	&& \left.  + \frac{4i\omega \epsilon (1-\epsilon)(1+\epsilon)^2}{ (1-\xi)^{\frac{2+\epsilon}{1+\epsilon}}}
	- \frac{2i\omega\epsilon(1-\epsilon)(1+\epsilon)^2}{ (1-\xi)^{2+\frac{1}{1+\epsilon}}}
	-i\omega \epsilon^2 (1+\epsilon)^2       
	- \frac{i\omega(1-\epsilon^2)^2}{ (1-\xi)^{\frac{2}{1+\epsilon}}}  \right.  \nonumber \\
	&& \left. - \frac{i\omega(1-\epsilon^2)^2}{ (1-\xi)^{\frac{2(2+\epsilon)}{1+\epsilon}}}
	+ \frac{2i\omega(1-\epsilon^2)^2}{ (1-\xi)^{\frac{3+\epsilon}{1+\epsilon}}}
	-\frac{\epsilon(1-\epsilon^2)[3+\epsilon-2i\omega(3-\epsilon)]}{ (1-\xi)^{2-\frac{1}{1+\epsilon}}}  \right.  \nonumber \\
	&&\left.  +\frac{\epsilon(1-\epsilon^2)[1-2i\omega(1-\epsilon)]}{ (1-\xi)^{3-\frac{1}{1+\epsilon}}} 
	+\frac{(1-\epsilon^2)^2-4i\omega +2i\omega \epsilon(2+3\epsilon+\epsilon^3)}{1-\xi}  \right.  \nonumber \\
	&&  \left. -\frac{(1-\epsilon^2)^2-i\omega (1+\epsilon)[6-\epsilon(14-9\epsilon+3\epsilon^2)]}{(1-\xi)^2}  \right\} ~.
\end{eqnarray}

We now expand $\lambda_0$ and $s_0$ in a Taylor series around a point $\xi_0$.  After selecting an appropriate depth [$n$ in Eq.\ (\ref{recurrenceforcoeffs})], we substitute the coefficients into ($\ref{recurrenceforcoeffs}$) to obtain an equation in $\omega$.  A root finder is then used to find the QNMs.  Although the choice of $\xi_0$ should not make a difference, in practice there are small variations when changing $\xi_0$.  We find setting $\xi_0$ to the $\xi$-value corresponding to the  maximum of the potential produces the best results.


We present the results for $l=0,1,2,6$  for different values of the parameter $\epsilon$ in Table IV. These results are in good agreement with the WKB results presented in Table I. The roots produced by the AIM become more accurate as the depth ($n$ in Eq.\ (\ref{recurrenceforcoeffs})) increases.  The roots included in Table IV are calculated using a depth of $110$.  However, to determine which of the roots ($\omega_n$) found at this depth are most reliable, we compare them with roots calculated at a depth of 100 ($\omega'_n$) and throw out those where  $|\omega_n-\omega'_n|<0.01$.  For higher values of $l$ and lower values of $\epsilon$, the AIM is able to find more roots for this particular black hole.

\vspace{0.2cm}
\footnotesize
\begin{tabular}{cccccc}
	\multicolumn{6}{c}{Table IV:  $\omega$ for $V_{IA}$ for different values of $\epsilon$ using AIM ~~~~~~~~~~~~~} \\ 
	\hline
	$n,l$ & $\epsilon=0$ &  $\epsilon=0.01$ & $\epsilon=0.1$    &\\ 
	\hline 
	0,0 & $0.22091-0.20979i$ &  $0.22212 - 0.20937 i$ & $ 0.23333 - 0.20557 i$  \\ 
	\hline 
	0,1 & $0.58587-0.19532i$ &  $0.58708 - 0.19514i$ & $0.59773 - 0.19313 i$  \\ 
	1,1 & $0.52890-0.61251i$ &  $0.52992 - 0.61196i$ & $0.53867 - 0.60543i$   \\ 
	2,1 & $0.45908-1.0803i$ &  $0.45913 - 1.0817i$ & $                    $  \\
	\hline 
	0,2 & $0.96729-0.19352i$ &  $0.96890 - 0.19336i$ & $0.98272 - 0.19150 i$  \\ 
	1,2 & $0.92770-0.59121i$ &  $0.92922 - 0.59070i$ & $0.94205 - 0.58496 i$   \\ 
	2,2 & $0.86109-1.0171i$ &  $0.86235 - 1.0163i$ & $0.87256- 1.0067i$ \\
	3,2 & $0.78773-1.4762i$ &  $0.78806 - 1.4754 i$ & $                $ \\
	\hline 
	0,6 & $2.5038-0.19261i$ &  $2.5074 - 0.19246 i$ & $2.5383 - 0.19067  i$  \\ 
	1,6 & $2.4875-0.57947i$ &  $2.4911 - 0.57900i$ & $2.5217 - 0.57360i$   \\ 
	2,6 & $2.4557-0.97120 i$ &  $2.4593 - 0.97041 i$ & $2.4891 - 0.96130  i$  \\
	3,6 & $2.4098-1.3708i$ &  $2.4133 - 1.3697 i$ & $2.4418 - 1.3568i$  \\
	4,6 & $2.3521-1.7810i$ &  $2.3554 - 1.7795  i$ & $2.3820 - 1.7628i$  \\
	5,6 & $2.2854-2.2035i$ &  $2.2884 - 2.2018i$ & $2.3124 - 2.1815i$  \\
	6,6 & $2.2130-2.6396i$ &  $2.2157 - 2.6375i$ & $2.2365 - 2.6142 i$  \\
	7,6 & $2.1383 - 3.0889i$ &  $2.1405 - 3.0867 i$ & $$  \\
	8,6 & $2.0641 - 3.5506i$ &  $2.0683 - 3.5500 i$ & $$  \\
	\label{Table1}
\end{tabular} 
\normalsize



\def\jnl#1#2#3#4{{#1}{\bf #2} #3 (#4)}

\def\Zphys{{Z.\ Phys.} }
\def\jssc{{J.\ Solid State Chem.\ }}
\def\jpsJ{{J.\ Phys.\ Soc.\ Japan }}
\def\ptps{{Prog.\ Theoret.\ Phys.\ Suppl.\ }}
\def\PTP{{Prog.\ Theoret.\ Phys.\  }}
\def\LNC{{Lett.\ Nuovo.\ Cim.\  }}

\def\JMP{{J. Math.\ Phys.} }
\def\NPB{{Nucl.\ Phys.} B}
\def\NP{{Nucl.\ Phys.} }
\def\PLB{{Phys.\ Lett.} B}
\def\PL{{Phys.\ Lett.} }
\def\PRL{Phys.\ Rev.\ Lett.\ }
\def\PRA{{Phys.\ Rev.} A}
\def\PRB{{Phys.\ Rev.} B}
\def\PRD{{Phys.\ Rev.} D}
\def\PR{{Phys.\ Rev.} }
\def\PRe{{Phys.\ Rep.} }
\def\AP{{Ann.\ Phys.\ (N.Y.)} }
\def\RMP{{Rev.\ Mod.\ Phys.} }
\def\ZPC{{Z.\ Phys.} C}
\def\SCI{Science}
\def\CMP{Comm.\ Math.\ Phys. }
\def\MPLA{{Mod.\ Phys.\ Lett.} A}
\def\IJMPA{{Int.\ J.\ Mod.\ Phys.} A}
\def\IJMPB{{Int.\ J.\ Mod.\ Phys.} B}
\def\cmp{{Com.\ Math.\ Phys.}}
\def\JPA{{J.\  Phys.} A}
\def\CQG{Class.\ Quant.\ Grav.~}
\def\ATMP{Adv.\ Theoret.\ Math.\ Phys.~}
\def\AJP{Am.\ J.\ Phys.~}
\def\PRSA{{Proc.\ Roy.\ Soc.\ Lond.} A }
\def\ibid{{ibid.} }
\vskip 1cm

\leftline{\bf References}

\renewenvironment{thebibliography}[1]
        {\begin{list}{[$\,$\arabic{enumi}$\,$]}  
        {\usecounter{enumi}\setlength{\parsep}{0pt}
         \setlength{\itemsep}{0pt}  \renewcommand{\baselinestretch}{1.2}
         \settowidth
        {\labelwidth}{#1 ~ ~}\sloppy}}{\end{list}}



\begin{thebibliography}{99}
\small
\baselineskip=16pt

\bibitem{Penrose1965} R.~Penrose, Gravitational Collapse and Space-Time Singularities, \jnl{\PRL}{14}{57}{1965}.

\bibitem{InfoLoss} See the review by D.~Harlow,  Jerusalem Lectures on Black Holes and Quantum Information, \jnl{Rev. Mod. Phys.}{88}{015002}{2016}

\bibitem{Page1993} D.~Page, Information in Black Hole Radiation, \jnl{\PRL}{71}{3743}{1993}.

\bibitem{reg9}K.A.\ Bronnikov, V.N.\ Melnikov and H.\ Dehnen, On a general class of brane-world black holes, \jnl{\PRD}{68}{024025}{2003}.


\bibitem{Teukolsky}M. Okounkova, L.C. Stein, M.A. Scheel, and S.A. Teukolsky, Numerical binary black hole collisions in dynamical Chern-Simons gravity, \jnl{\PRD}{100}{104026}{2019}.

\bibitem{reg1} J.M.\ Bardeen, Non-singular general-relativistic gravitational collapse, in Proceedings of the International Conference GR5, Tbilisi, USSR (Tbilisi University Press,
Tbilisi, 1968), p.\ 174.

\bibitem{PoissonIsrael} E.~Poisson and W.~Israel, Structure of the Black Hole Nucleus, \jnl{\CQG}{5}{L201}{1988}.

\bibitem{reg2}I.\ Dymnikova, Vacuum nonsingular black hole,  {Gen.\ Rel.\ Grav.} {\bf 24} 235–242 (1992).

\bibitem{reg3}C.\ Barrabes and V.P.\ Frolov, How many new worlds are inside a black hole?, \jnl{\PRD}{53}{3215}{1996}.




\bibitem{reg4}M.\ Mars, M.M.\ Mart\'{i}n–Prats, J.M.M.\ Senovilla, Models of regular Schwarzschild black holes satisfying weak energy conditions, \jnl{\CQG}{13}{L51}{1996}.

\bibitem{reg5}A. Cabo and E. Ayon-Beato, About black holes without trapping interior, Int.~J.~Mod.~Phys.~A {\bf 14} 2013 (1999).


\bibitem{reg6}A.\ Bogojevic and D.\ Stojkovic, A Nonsingular black hole, \jnl{\PRD}{61}{084011}{2000}.

\bibitem{reg7}R.\ Casadio, A.\ Fabbri and L.\ Mazzacurati, New black holes in the brane-world?, \jnl{\PRD}{65}{084040}{2002}.









\bibitem{reg10}S.A.\ Hayward, Formation and evaporation of regular black holes, \jnl{\PRL}{96}{031103}{2006}.

\bibitem{reg11}K.A.\ Bronnikov and J.C.\ Fabris, Regular Phantom Black Holes, \jnl{\PRL}{96}{251101}{2006}.

\bibitem{reg12}A.\ Simpson and M.\ Visser,  Black-bounce to traversable wormhole, {JCAP} {\bf 02} 042 (2019). 




\bibitem{reg14}D.\ Glavan and C.\ Lin, Einstein-Gauss-Bonnet gravity in 4-dimensional space-time, \jnl{\PRL}{124}{081301}{2020}.


\bibitem{pk09} A.~Peltola and G.~Kunstatter, Complete, Single-Horizon Quantum Corrected Black Hole Spacetime, \jnl{\PRD}{79}{061501}{2009}; Effective Polymer Dynamics of D-Dimensional Black Hole Interiors, \jnl{\PRD}{80}{044031}{2009}.

\bibitem{modesto06}L.~Modesto, Loop Quantum Black Hole, \jnl{\CQG}{23}{5587}{2006};
Black Hole Interior from Loop Quantum Gravity,
{Adv.\ High Energy Phys.}~\textbf{2008}, 1 (2008); 
Space-Time Structure of Loop Quantum Black Hole, {Int.\ J.\ Theor.\ Phys.} {\bf 49} 1649 (2010).


\bibitem{Ashtekar1} A.~Ashtekar, J. Olmedo, and P. Singh, Quantum Transfiguration of Kruskal Black Holes,
\jnl{\PRL}{121}{241301}{2018}.
\bibitem{Ashtekar2}  A. Ashtekar, J. Olmedo, and P. Singh, Quantum Extension of the Kruskal Space-time, \jnl{\PRD}{98}{126003}{2018}.

\bibitem{DGMK}R.G. Daghigh, M.D. Green, J.C. Morey, and G. Kunstatter,  Scalar Perturbations of a Single-Horizon Regular Black Hole, \jnl{\PRD}{102}{104040}{2020}.

\bibitem{RW}T. Regge and J.A. Wheeler, Stability of a Schwarzschild Singularity, Phys. Rev. {\bf 108} 1063 (1957).






\bibitem{QNMLQGbh0}J.H. Chen and Y.J. Wang, “Complex frequencies of a massless scalar field in loop quantum
black hole spacetime,” {Chin. Phys.} B{\bf 20}  030401 (2011).


\bibitem{QNMreg1}S.\ Fernando and J.\ Correa, Quasinormal Modes of Bardeen Black Hole: Scalar Perturbations, \jnl{\PRD}{86}{064039}{2012}.

\bibitem{QNMreg2}K.A.\ Bronnikov, R.A.\ Konoplya, A.\ Zhidenko, Instabilities of wormholes and regular black holes supported by a phantom scalar field, \jnl{\PRD}{86}{024028}{2012}.

\bibitem{QNMreg3} A.\ Flachi, J. P. S. Lemos, Quasinormal modes of regular black holes, \jnl{\PRD}{87}{024034}{2013}.

\bibitem{QNMreg4}K.\ Lin, J.\ Li, S.Z.\ Yang, Quasinormal modes of Hayward regular black hole, {Int. J. Theor. Phys.} {\bf 52} 3771-3778 (2013).

\bibitem{QNMreg5}J. Li, M. Hong and K. Lin, Dirac quasinormal modes in spherically symmetric regular black holes, \jnl{\PRD}{88}{064001}{2013}.

\bibitem{QNMreg6}S. Fernando, T. Clark, Black holes in massive gravity: quasi-normal modes of scalar perturbations, {Gen.\ Relativ.\ Gravit.} {\bf 46} 1834 (2014).


\bibitem{QNMreg7}C.F.B.\ Macedo, L.C.B.\ Crispino, Absorption of planar massless scalar waves by Bardeen regular black holes, \jnl{\PRD}{90}{064001}{2014}.



\bibitem{QNMreg8}J.\ Li, K.\ Lin, N.\ Yang, Nonlinear electromagnetic quasinormal modes and Hawking radiation of a regular black hole with magnetic charge, {Eur. Phys. J.} C{\bf 75} 131 (2015).

\bibitem{QNMreg9}B. Toshmatov, A. Abdujabbarov, Z. Stuchlk and B. Ahmedov, Quasinormal modes of test fields around regular black holes, \jnl{\PRD}{91}{083008}{2015}.

\bibitem{QNMLQGbh1}V. Santos, R.V. Maluf, C.A.S. Almeida, Quasinormal frequencies of self-dual black holes, \jnl{\PRD}{93}{084047}{2016}.


\bibitem{QNMreg10}J. Li, K. Lin, H. Wen, W.-L. Qian,  Gravitational Quasinormal Modes of Regular Phantom Black Hole, {Advances in High Energy Phys.~}2017  5234214 (2017).

\bibitem{QNMreg11}B. Toshmatov, Z. Stuchlk, J. Schee and B. Ahmedov, Electromagnetic perturbations of black holes in general relativity coupled to nonlinear electrodynamics, \jnl{\PRD}{97}{084058}{2018}.

\bibitem{QNMreg12}B. Toshmatov, Z. Stuchlk and B. Ahmedov, Electromagnetic perturbations of black holes in general relativity coupled to nonlinear electrodynamics: Polar perturbations, \jnl{\PRD}{98}{085021}{2018}.

\bibitem{QNMreg13}B. Toshmatov, Z. Stuchlk, B. Ahmedov, D. Malafarina, Relaxations of perturbations of spacetimes in general relativity coupled to nonlinear electrodynamics, \jnl{\PRD}{99}{064043}{2019}.

\bibitem{QNMreg14}H. Chakrabarty, A. A. Abdujabbarov and C. Bambi, Scalar perturbations and quasi-normal modes of a non-linear magnetic-charged black hole surrounded by quintessence, {Eur. Phys. J.~}C{\bf 79} 179 (2019).



\bibitem{QNMLQGbh2}M.B. Cruz, C.A.S. Silva, F.A. Brito,  Gravitational axial perturbations and quasinormal modes of loop quantum black holes, {Eur. Phys. J.} C{\bf 79} 157 (2019).


\bibitem{QNMreg15} Liu, Tao Zhu, Qiang Wu, Kimet Jusufi, Mubasher Jamil, Mustapha Azreg-Aïnou, Anzhong Wang, Shadow and Quasinormal Modes of a Rotating Loop Quantum Black Hole, \jnl{\PRD}{101}{084001}{2020}.

\bibitem{Ashtekar3}  A. Ashtekar and J. Olmedo, Properties of a recent quantum extension of the Kruskal geometry, Int.J.Mod.Phys.D {\bf 29} (2020) 10, 2050076; arXiv:2005.02309.


\bibitem{Bouhmadi}M. Bouhmadi-López, S. Brahma, C.-Y. Chen, P. Chen, D. Yeom, Asymptotic non-flatness of an effective black hole model based on loop quantum gravity, Phys.Dark Univ. {\bf 30} (2020) 100701


\bibitem{Faraoni} V.\ Faraoni, A.\ Giusti, Unsettling physics in the quantum-corrected Schwarzschild black hole, Symmetry 12 (2020) 1264, arXiv:2006.12577.

\bibitem{WKB1}B.F. Schutz and C.M. Will, Black hole normal modes - A semianalytic approach, {Astrophys. J.~} {\bf 291} L33 (1985).

\bibitem{WKB2}S. Iyer and C.M. Will, Black-hole normal modes: A WKB approach. I. Foundations and application
of a higher-order WKB analysis of potential-barrier scattering, \jnl{\PRD}{35}{3621}{1987}.



\bibitem{WKB-Konoplya}R.A. Konoplya, Quasinormal behavior of the D-dimensional Schwarzshild black hole and higher order WKB approach, \jnl{\PRD}{68}{024018}{2003}.

\bibitem{KonoplyaCode}These results are generated using {\em Mathematica} code provided by Roman Konoplya.




\bibitem{Price}R.H. Price,  Nonspherical Perturbations of Relativistic Gravitational Collapse.
I. Scalar and Gravitational Perturbations, \jnl{\PRD}{5}{2419}{1972}.

\bibitem{Prony}G. de Prony, Journal de l’\'{E}cole Polytechnique {\bf 1}(2), 24 (1795); S. L. Marple, Digital spectral analysis with applications, (Prentice-Hall, New Jersey, 1987).


\bibitem{AIM} H.T. Cho, A.S. Cornell, J. Doukas, T.R. Huang, W. Naylor, A New Approach to Black Hole Quasinormal Modes: A Review of the Asymptotic Iteration Method, Adv.~Math.~Phys.~2012 281705 (2012)




\end{thebibliography}
\end{document}